\documentclass[a4paper,10pt]{article}
\usepackage{amsmath,amssymb,graphicx,float}
\begin{document}

\title{Gauge Theories on de Sitter Spacetime}
\author{Asloob Ahmad Rather$^1$, Ashaq Hussain Sofi$^2$ \\
Shabir Ahmad Akhoon$^2$, Anil Maini$^3$ \\
$^1$Department of Physics, Aligarh Muslim University \\
U.P-202002, India \\
$^2$Department of Physics, National Institute of Technology \\
Srinagar, Kashmir-190006, India \\
$^3$Department of Applied Sciences, \\College of Engineering and Technology \\
BGSB University, Rajouri-185131, India}

\maketitle

\begin{abstract}
In this paper we will analyse the inner product for gauge theories in de Sitter spacetime. 
This will be done by   analysing an Euclidean version of the de Sitter metric. 
Thus, the de Sitter metric will be related to the metric on a  four-sphere. 
Then scalar spherical harmonics and vector spherical harmonics on a four-sphere will be analysed. 
An inner product for the gauge fields will be constructed using these modes. 
Finally, a two point function will be constructed, for gauge fields on de Sitter spacetime. 
\end{abstract}

It is known that all the fundamental forces in nature are described by 
gauge fields. Gauge fields are fields that are invariant under a gauge group. 
The electromagnetism is actually gauge theory generated by an abelian gauge  group 
called $U(1)$. The weak force is   generated by a non-abelian gauge group called 
$SU(2)$. Finally, the strong force is generated by a non-abelian gauge group called 
$SU(3)$. This all the three forces in nature can be described by gauge groups. 
In fact, we can argued that even we can regard gravity as a gauge theory of diffeomorphism. 
So, all the forces of nature can be regarded as gauge theories \cite{a}-\cite{b}. 
However, as all the degrees of freedom of a   gauge theories are not physical, 
it cannot be quantized directly. If we try to quantize a gauge theory, without fixing a 
gauge, it will lead to divergences in the generating functional of the Feynman graphs. 
Thus, we need to fix a gauge in order to quantize such a gauge theory. This can be done 
by adding a gauge fixing term to the original action. Furthermore, in order to keep the theory
unitary, we also need to add a ghost term to the original action. Thus, after adding a gauge fixing term and 
a ghost term a gauge theory can be quantized
\cite{e}-\cite{smu}.
It may be noted that gravity is not renormalizable. 
However, it can be argued that gravity can still be quantized in the framework of 
effective field theories. This is because there is no fundamental difference between 
renormalizable and non-renormalizable theories, in effective field theories, except the way 
the theory depends on
  lower 
energy scale. Now for a non-renormalizable theories, we will keep obtaining new 
constants at at every energy scale. However, once these constants have been fixed, 
at a particular energy scale by experiments, we can calculate any other perturbative 
diagram at that energy scale. So, even though for perturbative quantum gravity, 
we will get new constants at every energy scale, we can in principle fix them, 
by experiments at a energy scale. Then we can calculate any phenomena at that energy scale, 
using  perturbative quantum gravity. 
It may be noted that in the inflationary era, universe can be approximated by  
de Sitter spacetime. Furthermore, it is expected that our universe may be 
asymptotically approaching de Sitter spacetime,. So, we need to analyse the quantum field theory 
on de Sitter spacetime. 
In fact, quantum field theory on curved spacetime is interesting in its own right. 
\cite{7a}-\cite{wd1}. 
This is because both the Hawking radiation   is derived by using 
 quantum field theory in curved spacetime. 
 In this paper, we will study the inner product used in quantum field theories   for de Sitter spacetime. 
The de   Sitter  spacetime is  a spacetime of constant positive curvature. The positive curvature is generated by a 
 cosmological constant, which  sets the rate of expansion. The rate of expansion is measured by the  Hubble's constant  $H$. 
 So, we can relate the  Hubble's constant  $H$, with the  cosmological constant $\lambda$ as 
\begin{equation}
 \lambda = 3 H^2.
\end{equation}
Furthermore, this Hubble's constant is given by   the inverse of the radius of  de Sitter spacetime $r$, 
\begin{equation}
r = \frac{1}{H}.
\end{equation}
The de Sitter  spacetimes is a spacetime of constant curvature. In this spacetime, the curvature tensor can be expressed as,  
\begin{equation}
 R_{abcd} = \frac{1}{12}R[g_{ac}g_{bd}-g_{ad}g_{bc}].
\end{equation}
We can also write the following equation for de Sitter spacetime, 
\begin{eqnarray}
 R_{bd} &=& g^{ac} R_{abcd}  \\ &=& \frac{1}{12}R g^{ac} [g_{ac}g_{bd}-g_{ad}g_{bc}]\nonumber \\ &=& \frac{1}{4}R g_{db}.
\end{eqnarray}
Using these equations, we can express the Einstein tensor $G_{ab}$ in de Sitter spacetime as 
\begin{eqnarray}
 G_{ab} &=& R_{ab} - \frac{1}{2}R g_{ab} \nonumber \\ 
&=& \frac{1}{4}g_{ab} R - \frac{1}{2}R g_{ab} \nonumber \\
&=& - \frac{1}{4}g_{ab} R.
\end{eqnarray}
This equation can be viewed as a  solution to the Einstein equations, in  empty space with a  cosmological constant $\lambda$ , 
\begin{equation}
 \lambda = \frac{1}{4}R.
\end{equation}
Now we can also write these equations using the Hubble's constant, 
\begin{eqnarray}
 R_{abcd}& = &H^2 [g_{ac} g_{bd} - g_{ad} g_{bc} ],\nonumber\\
 R_{ab} &=& 3 H^2 g_{ab},\nonumber\\
 R &=& g^{ab}R_{ab} = 12 H^2,\nonumber\\
 G_{ab} &=& - 3 H^2 g_{ab} \nonumber\\
\end{eqnarray}

We can write the de Sitter metric as 
  \begin{equation}
  ds^2 =  -dt^2 + r^2 \cosh^2 r^{-1}t [d\chi^2 + \sin^2 \chi (d^2\theta + \sin^2 \theta d \phi )]. 
 \end{equation}
It may be noted that singularities at  $\chi =0, \pi$ and $\theta = 0, \pi$ 
 occur usually in polar coordinates. 
Apart from  these singularities, it      cover all of de Sitter spacetime. 
We can also write this metric as 
\begin{equation}
ds^2 = r^2 [-dt^2 + \cosh^2 t d\Omega].
\end{equation}
where the metric on a three-sphere is denoted by $d\Omega$. 
If we define $\Phi$ as
\begin{equation}
 \Phi = \frac{\pi}{2} -it.
\end{equation}
Then this becomes the  metric  on the four-sphere,
\begin{equation}
 ds_4^2 = r^2 [d\Phi^2 + \sin \Phi^2 d\Omega].
\end{equation}
 Now we can define vector spherical harmonics on $S^4$ as, 
\begin{eqnarray}
 -\nabla^2_n A_a &=& [L(L+3)-1] A_a, \nonumber \\
\nabla^a A_a &=& 0. 
\end{eqnarray}
There are two kinds of solutions to these equations, which are denoted by  $A^n$ where $n = 0, 1$. 
They can be written as 
\begin{eqnarray}
 A^1_\chi &=& 0 , \\ A^1_i& =& n_1 P_{L+1} Y_i, \\
A^0_\chi &=& n_2 (\sin\chi)^{-2} P_{L+1}Y, \\ A^0_i &=& n_2  \frac{1}{\ell(\ell+2)} D_1 P_{L+1}\nabla_i Y. 
\end{eqnarray}
Here 
the normalization constants $n_1$ and $n_2$ are  given by, 
\begin{equation}
 \int d^4 x \sqrt{g} g^{ab} A_a^{L} A_b^{*L'} = \delta^{LL'}. \label{nv}
\end{equation}
Here the scalar harmonics are denoted by $Y  $, which is a short hand notation for $ Y_{Llpm}$. 
The  associated Legendre function $P^{-l}_{L}(x)$ can be written as 
\begin{equation}
P^{-l}_{L}(x) = \frac{1}{\Gamma(1+l)}\left(\frac{1-x}{1+x}\right)^{\frac{l}{2}} F(-L, L+1, l+1, \frac{1-x}{2}).
\end{equation}
Here  $\Gamma(1+l) $ is the Gamma function,  $F(L, L+1, l+1, (1-x)/2)$ is the hypergeometric function.
The hypergeometric function  $F(a,b,c,x)$ in general is given by
 \begin{equation}
  F(a,b,c,x) = 1 + \frac{ab}{c}x + \frac{a(a+1)b(b+1)}{2! c(c+1)} x^2 + \cdots.
 \end{equation}
We also have 
\begin{equation}
 \left(  (1-x^2)\frac{d}{dx} +L x\right)P^{-l}_{L}(x) = (L-l)P_{L-1}^{-l}(x), 
\end{equation}
and 
\begin{equation}
 \left(  (1-x^2)\frac{d}{dx} -(L+1) x\right)P^{-l}_{L}(x) = -(L+l+1)P_{L+1}^{-l}(x).
\end{equation}
We can also define
\begin{equation}
 D_m = \frac{d}{d\chi} + m\cot\chi.
\end{equation}
So, we can write 
\begin{eqnarray}
  \left[ \frac{d}{d\chi} + m\cot\chi \right] (\sin\chi)^n f(\chi) = && \nonumber \\
  \sin^n\chi \left[ \frac{d}{d\chi} + (m+n)\cot\chi \right] f(\chi),&&
\end{eqnarray}
and 
\begin{equation}
 D_m \sin^n \chi f(\chi) = \sin^n \chi D_{m+n} f(\chi).
\end{equation}
Finally, we have 
\begin{equation}
 -\sin \chi D_n = \left[ (1- \cos\chi^2)\frac{d}{d\cos\chi} - n \cos\chi \right]. 
\end{equation}
Thus, we get 
\begin{eqnarray}
   -\sin \chi D_{-L} P^{-l}_{L}(\cos\chi) &=& (L-l)P_{L-1}^{-l}(\cos\chi),
\\
-\sin \chi D_{L+1} P^{-l}_{L}(\cos\chi) &=& -(L+l+1)P_{L+1}^{-l}(\cos\chi).
\end{eqnarray}
 The scalar spherical harmonics in one dimensions, can define as
\begin{equation}
 - \nabla^2_1 Y_m =  m^2 Y_m, 
\end{equation}
such that 
\begin{equation}
 Y_m =  \frac{1}{\sqrt{2\pi}} \exp(im\phi).
\end{equation}
It can also be defined in two dimensions as 
\begin{eqnarray}
 - \nabla^2_1 Y_{pm} &=&  m^2 Y_{pm}, \nonumber \\
- \nabla^2_2 Y_{pm} &=&  p(p+1) Y_{pm}, 
\end{eqnarray}
such that 
\begin{equation}
 Y_{pm}= c_2 P^{-m}_p Y_{m}.
\end{equation}
Furthermore, in three dimensions, we have 
\begin{eqnarray}
 - \nabla^2_1 Y_{lpm} &=&  m^2 Y_{lpm}, \nonumber \\
- \nabla^2_2 Y_{lpm} &=&  p(p+1) Y_{lpm},\nonumber \\
- \nabla^2_3 Y_{lpm} &=&  l(l+2) Y_{lpm}, 
\end{eqnarray}
such that 
\begin{equation}
 Y_{lpm}= c_3(\sin\chi)^{1/2} P^{-p-1/2}_{l + 1/2} Y_{pm}.
\end{equation}
So, in four dimensions, we have 
\begin{eqnarray}
 - \nabla^2_1 Y_{Llpm} &=&  m^2 Y_{Llpm}, \nonumber \\
- \nabla^2_2 Y_{Llpm} &=&  p(p+1) Y_{Llpm},\nonumber \\
- \nabla^2_3 Y_{Llpm} &=&  l(l+2) Y_{Llpm}, \nonumber \\
- \nabla^2_4 Y_{Llpm} &=&  L(L+3) Y_{Llpm}.
\end{eqnarray}
such that 
\begin{equation}
 Y_{Llpm}= c_4\sin\chi P^{-l-1}_{L + 1} Y_{lpm}.
\end{equation}
Here $c_4$ is a  normalization constant, which is given by 
\begin{equation}
 \int d^4x \sqrt{g} Y_{Llpm}Y^*_{L'l'p'm'} = \delta_{LL'}\delta_{ll'}\delta_{pp'}\delta_{mm'}. \label{ns}
\end{equation}
This gives us 
\begin{equation}
  c_4 = \left[ \frac{(2L +3)(L+l+2)!}{2(L+l)!}\right]^{\frac{1}{2}}.
\end{equation}

Now let us take a gauge field, $A_{\mu}^A T_A= A_\mu $. Here the $T_A$ are generators of a Lie algebra 
$[T_A, T_B] = i f_{AB}^C T_C$. 
Now we will write the Lagrangian for this theory as, 
\begin{equation}
 \mathcal{L} = -\sqrt{-g}[\mathcal{L}_1  + \frac{\alpha}{2} \mathcal{L}_2],
\end{equation}
where $\mathcal{L}_1$ is the original Lagrangian for the gauge field. 
It is given by 
\begin{equation}
 \mathcal{L}_1 = -\frac{1}{4} Tr [F^{\mu\nu}F_{\mu\nu}], 
\end{equation}
where $F_{\mu\nu} = F_{\mu\nu}^A T_A$, and 
\begin{equation}
  F_{\mu\nu}^A = \nabla_\mu A_\nu^A - \nabla_\nu A^A_\mu + g f^A_{BC}A^B_\mu A^C_\nu.
\end{equation}
The sum of the gauge fixing term and the ghost term can be written as 
$\mathcal{L}_2$.
Furthermore, we can write the action $S$ as
 \begin{equation}
  S = \int d^4 x \sqrt{-g}\mathcal{L}.
 \end{equation}
The equations of motion can be derived from this action. 

Now we define, $\pi^{\mu \nu}$ as the momentum current for this gauge field, 
 \begin{equation}
  \pi^{\mu \nu} = \frac{1}{\sqrt{-g}}\frac{\partial \mathcal{L}}{\partial \nabla_\nu A_\mu }.
 \end{equation}
Now   the two solutions to the field equations 
are written as $A_{\mu1}$ and $A_{\mu 2}$. 
We can define a current using these two solutions $J^c_{(A_{1}, A_{2})}$ as, 
 \begin{equation}
  J^\nu = i [A_{\mu1}^* \pi^{\mu\nu}_2 - A_{\nu2} \pi^{*\mu\nu}_1].
 \end{equation}
This current is conserved
\begin{equation}
 \nabla_c J^c = 0.
\end{equation}
Now an inner product on a space-like hyper-surface $\Sigma_\nu$ can defined as  
 \begin{equation}
  (A^\mu_{1}, A_{\mu 2}) = \int d\Sigma_\nu J^\nu_{(A^\mu_{1}, A_{\nu2})}.
 \end{equation}
The inner product  can be simplified as follows, 
\begin{equation}
 (A^\mu_{1}, A_{\mu2}) =  \int d^3 x \sqrt{-g} J^0.
\end{equation}
This inner product will be conserved. 
Using a complete set of solutions to the classical equations,  $A_{\mu n}$ and $A^*_{ \mu n}$ , we can write 
 \begin{equation}
  A_\mu= \sum_n [a_{n} A_{\mu n} + a^*_{n} A^*_{ \mu n} ].
 \end{equation}
Furthermore, we can also write,  
\begin{equation}
 \pi^{\mu\nu} = \sum_n [a_n \pi^{\mu\nu}_n + a^*_n \pi^{*\mu\nu}_n ].
\end{equation}
Now we suppose that the following holds, 
 \begin{equation}
  (A^\mu_n,A^*_{\mu m} ) = 0
  \end{equation}
  and
\begin{equation}
    (A^\mu_n,A_{\mu m}) = M_{nm}.
\end{equation}
Now as $A_{\mu}$  is promoted to an operator, its modes also become operators. Thus, 
 $a_n $ become creation operators $a^{\dagger}_n$  
and  annihilation operators,  respectively.
Now we can write, 
 \begin{equation}
  \hat{A}_\mu = \sum_n [a_n A_{\mu n} + a^{\dagger}_n A^*_{\mu n} ].
 \end{equation}
So the  two-point function can be written as 
\begin{eqnarray}
 G_{II'} (x,x') &=& \langle 0| A_I(x) A_{I'}(x')|0\rangle \nonumber
 \\&=& \sum_{mn}A_{In}(x)A_{I' m}(x')\langle 0| [a_n, a_m^{\dagger}]|0\rangle.
\end{eqnarray}
Now we define  $C_{nm}$ as  
\begin{equation}
 C_{nm} = \langle 0|[a_n, a_m^{\dagger}]|0\rangle.
\end{equation}
So, we can write, 
\begin{equation}
 [(A_n, \hat{A})(\hat{A},  A_m)] = M_{nm}.
\end{equation}
We can also write,  
\begin{equation}
 M_{nm} = M^*_{mn}.
\end{equation}
This can be written in Matrix notation as 
\begin{equation}
 MCM = M.
\end{equation}
Thus, we get 
\begin{equation}
 C= M^{-1}.
\end{equation}
So, finally we can write, 
 \begin{equation}
   G(x,x')_{\mu\nu'} = \sum_{nm} A_{\mu n}A_{\nu'm} M^{-1}_{nm}.
 \end{equation}
 
In this paper, we analysed the inner product for gauge theories on de Sitter spacetime. 
This was done by first analysing spherical harmonics on $S^4$ and then mode expanding the solutions 
of the gauge field in terms of these modes. Thus, we obtained an explicit form for the two-point function. 
This two-point function can be used to do perturbative calculations.

\end{document}